\documentclass[12pt]{article} %,a4paper
\usepackage{graphicx}
\usepackage{amsmath}
\usepackage{amssymb}
\usepackage{mathtools}
\allowdisplaybreaks

\usepackage{utopia}
\usepackage[utopia]{mathdesign}
 \usepackage{cite}
 \usepackage{hyperref}
\usepackage[height=8.85in,width=6.45in]{geometry}
\numberwithin{equation}{section}
\newcommand{\bC}{\mathbb{C}}
\newcommand{\bR}{\mathbb{R}}
\newcommand{\bZ}{\mathbb{Z}}

\def\SU{\mathrm{SU}}
\def\SL{\mathrm{SL}}
\def\ff{\mathfrak{f}}
\def\fg{\mathfrak{g}}
\def\fh{\mathfrak{h}}

\def\fsu{\mathfrak{su}}
\def\fso{\mathfrak{so}}
\def\fsp{\mathfrak{sp}}
\def\fe{\mathfrak{e}}
\let\oldmid\mid
\def\mid{\,\oldmid\,}

\def\ord{\mathop{\mathrm{ord}}}
\def\gcd{\mathop{\mathrm{gcd}}}

\def\iia{\mathrm{IIA}}
\def\iib{\mathrm{IIB}}
\def\II{I\!I}
\def\III{I\!I\!I}
\def\IV{I\!V}

\begin{document}

\begin{titlepage}

\begin{flushright}
IPMU-15-0138\\
UT-15-30
\end{flushright}

\vskip 2cm

\begin{center}

{\Large\scshape On {\bfseries M} and {\bfseries F} theory's}

\vskip-.3em

\includegraphics[width=.7\textwidth]{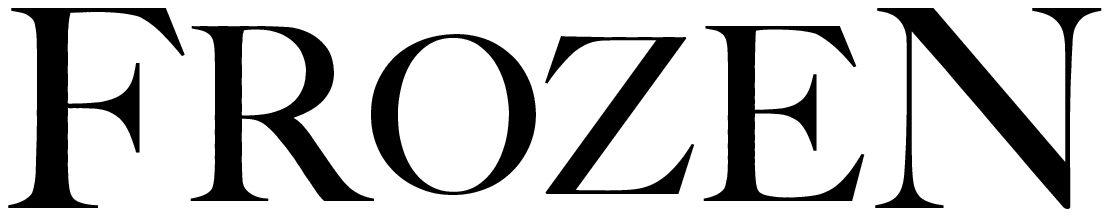}

\vskip-1em

{\Large\scshape singularities}

\vskip 1.5cm
 Yuji Tachikawa
\vskip 1.0cm

\begin{tabular}{ll}
  & Department of Physics, Faculty of Science, \\
& University of Tokyo,  Bunkyo-ku, Tokyo 113-0033, Japan, and\\
  & Institute for the Physics and Mathematics of the Universe, \\
& University of Tokyo,  Kashiwa, Chiba 277-8583, Japan\\
\end{tabular}

\vskip 1cm

\textbf{Abstract}

\end{center}

\medskip
\noindent
We revisit the duality between ALE singularities in M-theory and 7-branes on a circle in F-theory.  
We see that a frozen M-theory singularity maps to a circle compactification involving a rotation of the plane transverse to the 7-brane, showing an interesting correspondence between commuting triples in simply-laced groups and Kodaira's  classification of singular elliptic fibrations.
Our analysis strongly suggests that the O7$^+$ plane is the only completely frozen F-theory singularity.

\end{titlepage}

\section{Introduction and summary}
It is by now a textbook material that, in M-theory, a singularity locally of the form $\bC^2/\Gamma_\fg$ where $\Gamma_\fg$ is a finite subgroup of $\SU(2)$ corresponding to a simply-laced algebra $\fg$ produces a 7d super Yang-Mills theory with gauge algebra $\fg$ at the singular locus. 
It might be less-known that there are (partially) frozen variants of such a singularity,
still preserving 16 supercharges,
characterized by  a non-zero value of \begin{equation}
r:=\int_{S^3/\Gamma_\fg} C = \frac nd \mod 1 \label{r}
\end{equation} around it \cite{deBoer:2001px,Atiyah:2001qf,MorrisonTalk2002}.
Here, $C$ is the M-theory 3-form and $d$ is a label on the nodes of the Dynkin diagram of $\fg$ and $\gcd(n,d)=1$.
On such an M-theory singularity, we have a gauge algebra $\fh_r$.
We list possible values of $r$, $\fh_r$  for all $\fg$ in Table~\ref{M}.\footnote{Our convention is $\fsp(2)=\fsu(2)$. }

\begin{table}[h]
\[
\begin{array}{c@{\,}l||c|cc|ccc|ccccc}
&\fg &   \fso(2k+8)  &   \fe_6 & \fe_6 & \fe_7 &\fe_7 &\fe_7&  \fe_8& \fe_8& \fe_8& \fe_8& \fe_8 \\[1pt]
\hline
r=&\vphantom{\Big|} \frac nd &  \frac12 &    \frac12 & \frac13, \frac23 & \frac12 & \frac13,\frac23 & \frac14,\frac34 &   \frac12 & \frac13,\frac23 & \frac14,\frac34 & \frac15,\frac25,\frac35,\frac45 & \frac16,\frac56\\
&\fh_r &  \fsp(2k)  & \fsu(3) & * &  \fso(7) & \fsu(2) & * &  \ff_4 & \fg_2 & \fsu(2) & * & *\end{array}
\]
\caption{Partially frozen half-BPS M-theory singularities of the form $\bC^2/\Gamma_\fg$. \label{M}}
\end{table}

Let us now consider a situation where such a singularity is in a singular fiber of an elliptic fibration $X_g$ over the complex plane\footnote{In the following the subscript on $\bC$ denotes the symbol for its coordinate. } $\bC_z$,
such that there is an $\SL(2,\bZ)$ monodromy $g$ around $z=0$ acting on the elliptic fiber.
Possible conjugacy classes of monodromies were classified by Kodaira, see Table~\ref{F}.  
The central fiber can be appropriately blown up so that the whole geometry is smooth, but we mainly consider the case that there is a singularity there. 
When the monodromy is diagonalizable, the supergravity solution can be taken so that the base $\bC_z$ is  conical, with a metric of the form $d\!R^2+R^2d\!\theta^2$ with  a nontrivial periodicity $\theta\sim \theta+\theta_g$. We call $\theta_g$ the opening angle in this note. 

Now let us first consider standard singularities which are not (partially) frozen.
At each point on a base, we can reduce along an $S^1$ of the fiber and take the T-dual of the other $S^1$ to have a type IIB setup on $S^1$.
The monodromy around $z=0$ now acts on the axiodilaton of the type IIB theory, therefore we are now in an F-theory configuration: we have a 7-brane at $z=0$, compactified on $S^1$. 
As is well-known, when we shrink the elliptic fiber on the M-theory side, the $S^1$ on the F-theory side opens up.
We see that a 7-brane characterized by a monodromy $g$, obtained in a manner outlined above, has a gauge symmetry $\fg$ on it \cite{Morrison:1996na,Morrison:1996pp}.

\begin{table}[h]
\[
\begin{array}{c|ccc|ccc|l}
&g  & \theta_g/\frac\pi6 &  \fg & \ord(a) & \ord(b) & \ord(\Delta)  &\text{sample equation} \\
\hline
I_k &   \begin{pmatrix}
1 & k \\
0 & 1
\end{pmatrix} & 12-k& \fsu(k)   &   0 &  0 & k  & y^2=x^3-3x+2+z^k \\
\II  & \begin{pmatrix}
1 & 1 \\
-1 & 0
\end{pmatrix} & 10&  *  &  1 & 1 & 2 & y^2=x^3 + z \\
\III  & \begin{pmatrix}
0 & 1 \\
-1 & 0
\end{pmatrix} & 9& \fsu(2) & 1 &  2 & 3  &  y^2=x^3+zx \\
\IV  &  \begin{pmatrix}
0 & 1 \\
-1 & -1
\end{pmatrix} & 8&\fsu(3) & 2 & 2 &  4 & y^2=x^3+z^2 \\
I_k^* & \begin{pmatrix}
-1 & -k\\
0 & -1
\end{pmatrix} & 6-k&  \fso(2k+8) & 2 & 3 & k+6   & y^2 = x^3 -3 z^2 x + 2z^3(1+z^k) \\
 \IV ^*  &\begin{pmatrix}
-1 & -1 \\
1 & 0
\end{pmatrix} & 4& \fe_6  & 3 & 4 & 8 & y^2=x^3+z^4\\
\III ^* &\begin{pmatrix}
0 & -1 \\
1 & 0
\end{pmatrix} & 3& \fe_7 & 3 & 5 & 9  & y^2=x^3+z^3x\\
 \II ^* & \begin{pmatrix}
0 & -1 \\
1 & 1
\end{pmatrix} &   2& \fe_8 & 4 & 5 & 10  & y^2=x^3+z^5
\end{array}
\]
\caption{Possible monodromies $g$ of elliptic fibrations over the base $\bC_z$.  Here,  $\fg$ is the type of the singularity and $\theta_g$ is the opening angle.
The equation of the elliptic fiber is given by $y^2=x^3+a(z)x+b(z)$ where
 $z$ is the coordinate of the base,  and $\Delta=4a^3+27b^2$. 
 \label{F}}
\end{table}

It is known, however, that the monodromy alone does not completely characterize a 7-brane in F-theory.
For example, the singular fiber of type $I^*_4$ can be realized in perturbative type IIB string theory by putting 8 D7-branes\footnote{Without counting mirror images.} on top of an O7$^-$ plane\footnote{Our convention is that  O$^-$ planes give orthogonal symmetries and O$^+$ planes give symplectic symmetries.}, and also by just an O7$^+$ plane. 
The former has $\fso(16)$ gauge algebra on it but the latter does not have any, and therefore they are clearly distinct.
Correspondingly, the former $I^*_4$ singularity can be deformed, but the latter $I^*_4$ singularity is somehow completely frozen, probably due to an effect of a discrete flux \cite{Witten:1997bs}.

This begs a natural question: \emph{are there other (partially) frozen variants of F-theory 7-branes?}
Clearly $I^*_{4+k}$ singularities have two versions: one given by an O7$^-$ plane plus $8+k$ D7-branes,
another given by an O7$^+$ plane plus $k$ D7-branes.
The main objective of this note is to argue that \emph{there are no other (partially) frozen half-BPS 7-branes.} 

We approach this question by first studying an M-theory configuration on an elliptic fibration $X_g$ over $\bC_w$ with a singularity of type $\fg$ at the central fiber, now with a nonzero value of $r=\frac nd$ defined in \eqref{r}.
In Sec.~2, we will see that, a fiber-wise duality to the type IIB description, we have  an F-theory configuration on $(\bC_{z} \times S^1)/\bZ_d$ where $w=z^d$ with monodromy $g^d$ around $z=0$, such that the quotient is given by $z\mapsto e^{2\pi i n/d} z$ together with a $\frac1d$ shift along $S^1$.

We then run the arguments in reverse in Sec.~3. 
Namely, we take a putative (partially) frozen half-BPS 7-brane in F-theory, and compactify it on $S^1$. 
We will take a fiber-wise duality to go to an M-theory frame. 
This should result in an elliptic fibration with singularities at the origin of the base $\bC$.
Using the list of (partially) frozen half-BPS singularities in M-theory, we conclude that the O7$^+$ planes with D7-branes on top are the only (partially) frozen half-BPS 7-branes in F-theory. 
We conclude with a brief discussion in Sec.~4.

\section{Frozen singularities in M-theory and their F-theory duals}
Let us start by commenting further on the structure of the (partially) frozen singularities of the form $\bC^2/\Gamma_\fg$ in M-theory \cite{deBoer:2001px,Atiyah:2001qf}. 
As already recalled, they are characterized by a non-zero value of $
r=\int_{S^3/\Gamma_\fg} C \mod 1
$ around the singularity. 
The allowed choices of $r$ correspond to  the Chern-Simons invariants of  flat $G$ bundles on $T^3$, where $G$ is the connected and simply-connected group for $\fg$.
A flat $G$ bundle on $T^3$ corresponds to a commuting triple of elements of $G$ up to conjugacy, and therefore this information $r$ is often called a \emph{triple}.

On such an M-theory singularity with non-zero $r$, we have a 7d super Yang-Mills theory with gauge algebra $\fh_r$
which is given by the Langlands dual of the subalgebra of $\fg$ commuting with this flat bundle, or equivalently the commuting triple.%
\footnote{These facts can be understood via the fractionation of M5-branes  and its relation to instantons on $T^3\times \bR$, see e.g.~\cite{DelZotto:2014hpa} and Sec.~3.1 of \cite{Ohmori:2015pua}. Readable accounts on triples can be found in the last section of \cite{Witten:2000nv} and the last subsection of \cite{Atiyah:2001qf}.} 
If $\fh_r$ is empty the singularity is completely frozen. If $\fh_r$ is nonempty, the singularity is only partially frozen and can be deformed to a completely frozen one,
whose type  is a minimal one compatible with a given value of $r$.

For example, take $\fg=\fe_8$ and $r=\frac13$. The minimal algebra compatible with this value of $r$ is $\fe_6$, whose nontrivial commuting triple is in fact contained in $\ff_4$. The commutant of this $\ff_4$ is $\fg_2$, whose Langlands dual gives $\fh_r=\fg_2$. 
As another example, take $\fg=\fso(2k+8)$ and $r=\frac12$. The minimal algebra compatible with this $r$ is $\fso(8)$, whose nontrivial commuting triple is in fact contained in $\fso(7)$. Its commutant within $\fg$ is $\fso(2k+1)$, whose Langlands dual is $\fh_r=\fsp(2k)$.

Take now an elliptic fibration $X_g$ over $\bC_w$ with a singularity of type $\fg$ in the singular fiber at $w=0$. We remind the reader that $g$ stands for the $\SL(2,\bZ)$ monodromy at $w=0$.
We furthermore put a nonzero value of $r$ to (partially) freeze the singularity.
We would like to construct an F-theoretic dual description of this setup. 

It is easiest to start with the case of the $I_k^*$ fiber with a $D_{k+4}$ singularity, (partially) frozen with $r=\frac12$.
When we reduce it to type IIA, this becomes an O6$^+$ plane plus $k$ D6-branes \cite{Landsteiner:1997ei,Witten:1997bs}.
Since the monodromy preserves an $S^1$ of the elliptic fiber up to a multiplication by $-1$,
we can reduce the whole setup globally along this $S^1$ to a genuine type IIA configuration on $(\bC_z\times S^1_\iia)/\bZ_2$ 
with an O6$^+$ plane plus $k$ D6-branes  on one of the fixed points.
On the other fixed point, we should have an O6$^-$ plane, which is known to lift to a smooth configuration in M-theory.
Note that we have $w=z^2$, since the two points $\pm z$ on $\bC_z$ are identified by the orientifolding action.

Now, take the T-dual to obtain a type IIB configuration. 
We have a so-called shift-orientifold on $(\bC_z\times S^1_\iib)/\bZ_2$, where the orientifold action on $\bC_z$ is accompanied by a $\frac12$ shift of $S^1_\iib$ with $2k$ D7-branes on the locus $z=0$,  see e.g.~\cite{Hanany:2001iy}.
Note that there is $\fsu(2k)$ gauge algebra locally on the D7-branes, which is broken to $\fsp(2k)$ by the compactification on $S^1$ involving the orientifolding action, that acts as an outer automorphism of $\fsu(2k)$.

The orientifolding action $z\mapsto -z$ when we go along $S^1_\iib$ can also be understood as follows: 
we had $\int_{S^3/\Gamma_\fg} C=\frac12$ around the singularity in the M-theory description.
This 3-cycle can be deformed into a 3-cycle $T$ given by a large $S^1_\text{big} \subset \bC_z$ times the elliptic fiber.  
In the type IIA reduction, this means that there is $\int_{S^1_\text{big}\times S^1_\iia} B=\frac12$,
which turns into a $\frac12$ rotation in the type IIB setup. 

Let us summarize what we have found in an F-theoretic language.
We started from the $I_k^*$ fiber with monodromy $g=\begin{psmallmatrix}
-1 & -k \\
0 & -1
\end{psmallmatrix}$ on $\bC_w$, with $r=\frac12$ in M-theory. 
The result is an F-theory configuration on $\bC_{z=w^{1/2}}$ with the $I_{2k}$ fiber whose monodromy is $g^2=\begin{psmallmatrix}
1 & 2k\\
0 & 1
\end{psmallmatrix}$, further compactified on $S^1$ such that $z\mapsto -z$ when we go around $S^1$.
Now, the $I_{2k}$ singularity can be written as $y^2=x^2+z^{2k}$ and the action $z\mapsto -z$ preserves this form. Furthermore, this action is known to correspond to a $\bZ_2$ outer automorphism of the $\fsu(2k)$ gauge algebra of the 7-brane of type $I_{2k}$.

\begin{table}
\centerline{
$\begin{array}{c|c|c|c|c||c|c|c|c|c|c}
 g  & & \theta_g/\frac\pi6 & \text{alg.}  & r=\frac nd & g^d &  & \theta_{g^d}/\frac\pi6 & \text{alg.} & \text{outer} & \text{fixed}  \\
 \hline\hline
 \begin{pmatrix}
 -1 & -k \\
 0 & -1
\end{pmatrix} & I_k^* & 6-k & \fso(2k+8) & \frac12 &
\begin{pmatrix}
1 & 2k \\
0 & 1
\end{pmatrix} & I_{2k}& 12-2k & \fsu(2k) & \bZ_2 & \fsp(2k)\\
\hline
\begin{pmatrix}
 -1 & -1 \\
 1 & 0
\end{pmatrix} & \IV ^* &4& \fe_6  & \frac12 &
\begin{pmatrix}
0 & 1 \\
-1 & -1
\end{pmatrix} & \IV  & 8& \fsu(3)  & * & \fsu(3) \\
\begin{pmatrix}
 -1 & -1 \\
 1 & 0
\end{pmatrix} & \IV ^* & 4 & \fe_6 & \frac13,\frac23 &
\begin{pmatrix}
1 & 0 \\
0 & 1
\end{pmatrix} & I_0 & 12& *   & * & * \\
\hline
\begin{pmatrix}
 0 & -1 \\
 1 & 0
\end{pmatrix} & \III ^* &3 & \fe_7 & \frac12 &
\begin{pmatrix}
-1 & 0 \\
0 & -1
\end{pmatrix} & I_0^* &6& \fso(8) & \bZ_2 & \fso(7) \\
\begin{pmatrix}
 0 & -1 \\
 1 & 0
\end{pmatrix} & \III ^* & 3& \fe_7 & \frac13,\frac23 &
\begin{pmatrix}
0 & 1 \\
-1 & 0
\end{pmatrix} & \III  &9 & \fsu(2) & * & \fsu(2) \\
\begin{pmatrix}
 0 & -1 \\
 1 & 0
\end{pmatrix} & \III ^* & 3& \fe_7 & \frac14,\frac34 &
\begin{pmatrix}
1 & 0 \\
0 & 1
\end{pmatrix} & I_0 &12 &  * & * & * \\
\hline
\begin{pmatrix}
 0 & -1 \\
 1 & 1
\end{pmatrix} & \II ^* & 2 & \fe_8 & \frac12 &
\begin{pmatrix}
-1 & -1 \\
1 & 0
\end{pmatrix} & \IV ^* & 4 &  \fe_6 & \bZ_2 & \ff_4 \\
\begin{pmatrix}
0 & -1 \\
 1 & 1
\end{pmatrix} & \II ^* &2 &  \fe_8 & \frac13,\frac23 &
\begin{pmatrix}
-1 & 0 \\
0 & -1
\end{pmatrix} & I_0^* & 6& \fso(8) & \bZ_3 & \fg_2 \\
\begin{pmatrix}
0 & -1 \\
 1 & 1
 \end{pmatrix} & \II ^* & 2& \fe_8 & \frac14,\frac34 &
\begin{pmatrix}
0 & 1 \\
-1 & -1
\end{pmatrix} & \IV  & 8 & \fsu(3) & \bZ_2 &  \fsu(2) \\
\begin{pmatrix}
 0 & 1 \\
- 1 & -1
\end{pmatrix} & \II ^* & 2& \fe_8   & \frac15,\frac25,\frac35,\frac45 &
\begin{pmatrix}
1 & 1 \\
-1 & 0
\end{pmatrix} & \II  & 10 & * & * & * \\
\begin{pmatrix}
0 & -1 \\
 1 & 1
 \end{pmatrix} & \II ^* & 2 &  \fe_8 & \frac16,\frac56 &
\begin{pmatrix}
1 & 0 \\
0 & 1
\end{pmatrix} & I_0 &12 &  * & * & * 
\end{array}
$}
\caption{M-theory on an elliptic fibration with monodromy $g$ and a discrete flux $r=\frac nd$,
and its F-theory realization characterized by the monodromy $g^d$ and an outer-automorphism. 
\label{B}}
\end{table}

We find that the opening angles $\theta_{g}$ of $\bC_w$ and  $\theta_{g^d}$  of $\bC_z$  satisfy the relation $\theta_{g^d}=d\theta_g$, as it should be for the metric to be consistent.
The action $z\mapsto e^{2\pi i n/d} z$ % typo pointed out by S. Cherkis.
induces an outer automorphism of the algebra.
Exactly which automorphism is induced can in principle be determined from an F-theory computation, essentially the same one to determine the gauge algebra on non-split singularities.
Instead, here we used the information that the fixed subalgebra should be the gauge algebra $\mathfrak{h}_r$ on the M-theory singularity with the discrete flux $r$.

One consistency check is to compare the cases $g^d$ is of type $\IV$ and $d=2,4$. From the $d=4$ case, we see that the $2\pi/4$ rotation of $z$ corresponds to the generator of the $\bZ_2$ outer automorphism of $\fsu(3)$. This means that the $2\pi/2$ rotation should correspond to the trivial element of the $\bZ_2$ outer automorphism, which is indeed the case.

Before proceeding, we pause to mention that the rotation of the phase of $z$ by $2\pi \frac nd$ on the plane $\bC_z$ transverse to the 7-brane on the F-theory side can be derived by a further compactification on $S^1$.
Let us start from the M-theory side. 
We compactify the whole setup on $S^1$.
This is a type IIA configuration on an elliptic fibration $X_g$ with a non-zero $\int_T C_{(3)}=\frac nd$,
where, as before, $T$ is a big circle $S^1_\text{big}\subset \bC_w$ times the elliptic fiber, and $C_{(3)}$ is the RR 3-form. 
We then take a double T-dual along the elliptic fiber. 
This is again a type IIA configuration on an elliptic fibration with a non-zero $\int_{S^1_\text{big}} C_{(1)}=\frac nd$, where $C_{(1)}$ is now the RR 1-form\footnote{Such backgrounds were first considered in \cite{Schwarz:1995bj}.}.
This lifts to a new M-theory configuration on\[
\left(\text{(an elliptic fibration $X_{g^d}$ over $\bC_{z}$)} \times S^1 \right)/\bZ_d
\]
such that the generator of $\bZ_d$ is the rotation of the $z$ plane  by $2\pi \frac nd$ together with the $\frac1d$  shift of the M-theory circle.\footnote{On the one hand, close to the singularity in the singular fiber, this is essentially the configuration studied in \cite{Witten:1997kz}.  On the other hand, if we replace the elliptic fibration $X_g$ by a compact K3, such a background was first considered in \cite{Kachru:1997bz}.} Clearly this is the $S^1$ compactification of the F-theory configuration described above.

\section{Frozen F-theory 7-branes and their M-theory duals}
Let us now study (partially) frozen F-theory 7-branes,
 by running the argument of the previous section in reverse. 
Again, it is easiest to start with the case which has a perturbative type IIB realization. 

Let us consider an O7$^+$ plane.
It has a monodromy of type $I_4^*$. 
We compactify the whole system on $S^1$, and take the T-duality. 
We now have a type IIA system on $(\bC_w\times S^1)/\bZ_2$, with O6$^+$ planes on both fixed points.
Therefore, its M-theory lift has two frozen singularities of type $D_4$. 
Note that these two singularities are both on the same singular fiber of the elliptic fibration.
Since the sum of two Dynkin diagrams of type $D_4$ is contained in an affine Dynkin diagram of type $D_8$,
we see that the singular fiber has the type $I_4^*$, as it should be.

Note that each of the $D_4$ singularities has $\int_{T_a} C=\frac12 \mod 1$ around it, where $T_{a=1,2}$  is  the quotient of $S^3$ around each of the singularities. 
The three-cycle $T$ given by  $S^1_\text{big}\subset \bC_z$ times the elliptic fiber is their sum $T_1+T_2$, and therefore has $\int_T C= 0 \mod 1$. 
This is compatible with the fact that the $S^1$ compactification on the F-theory side does not involve any rotation. 

With this warm-up, let us consider a general (partially) frozen half-BPS 7-brane at $z=0$ of $\bC_z$, with monodromy $g$ around $z=0$.
We assume it preserves  16 supercharges.
Compactify it on $S^1$ and take the M-theory dual. 
This operation should be possible away from $z=0$ fiber-wise. 
We then have an M-theory configuration of an  elliptic fibration away from $z=0$, with the same monodromy $g$.
Given that it preserves 16 supercharges, it is strongly likely that the M-theory geometry is given by an elliptic fibration with singularities no worse than orbifolds of $\bC^2$ by finite subgroups of $\SU(2)$. 
Let us say there are $m$ singularities of type%\footnote{Here the second entry $\fg_2$ in the list does \emph{not} stand for an exceptional Lie algebra $\fg_2$.} 
$\fg_1$, \ldots, $\fg_m$ at the central fiber. 
At least two out of these $m$ singularities should be (partially) frozen;
otherwise we can change the K\"ahler parameter to have just zero or one frozen singularity, and we know those cases do not correspond to (partially) frozen 7-branes.

Therefore, at least two of $\fg_1$, \ldots, $\fg_m$ are of type $D$ or $E$, thus with three prongs. 
We also know that the sum of Dynkin diagrams of type $\fg_i$ is contained in an affine Dynkin diagram whose type is determined by the monodromy $g$. 
Now, by direct inspection, we can easily see that the only affine Dynkin diagrams that can contain more than one finite Dynkin diagrams with three prongs are of type $D_{8+k}$ with $k\ge 0$. 
The two finite Dynkin diagrams are necessarily of the type $D_{4+k_1}$ and $D_{4+k_2}$, with $k_{1,2}\ge 0$.
By our assumption both are (partially) frozen.

This is a type IIA configuration on $(\bC_w\times S^1)/\bZ_2$, with an O6$^+$ with $k_1$ D6-branes on one fixed point, and  an O6$^+$ with $k_2$ D6-branes on another.
Taking the T-dual, we have a type IIB configuration on $(\bC_w/\bZ_2) \times S^1 = \bC_z \times S^1$.
We have an O7$^+$ with $k_1+k_2$ D7-branes at $z=0$,
and the whole system is further compactified on $S^1$ with a Wilson line around it,
so that $\fsp(2k_1+2k_2)$ is broken to $\fsp(2k_1)\oplus \fsp(2k_2)$.
We conclude that \emph{a half-BPS (partially) frozen 7-brane is necessarily an O7$^+$ plane, possibly with an integral number of D7-branes on top.}  
Note that our analysis does \emph{not} allow a stuck $\frac12$  D7-brane on top of an O7$^+$, thus precluding the existence of half-BPS $\widetilde{\mathrm{O7}}^+$. 
This is consistent with the analysis in \cite{deBoer:2001px}.

\section{Discussions}
In this short note, we argued that there is a duality between 
\begin{itemize}
\item  M-theory configurations on an elliptic fibration on $\bC_w$ with monodromy $g$ around $w=0$ with a (partially) frozen singularity with $\int_{S^3/\Gamma_\fg} C=\frac nd \mod 1$, and 
\item  F-theory configurations of a 7-brane on $\bC_{z=w^{1/d}}$ with monodromy $g^d$ around $z=0$,
further compactified on $S^1$ so that $z$ is rotated as $z\mapsto e^{2\pi i n/d} z$ when we go around $S^1$.
\end{itemize}
We then argued that, using the same logic, a (partially) frozen half-BPS 7-brane is necessarily a combination of an O7$^+$ plane plus an integral number of D7-branes. 

Note that in our argument,  we assumed that Table~\ref{M} exhausted the list of (partially) frozen half-BPS codimension-4 singularities of M-theory.  Therefore, we can state our conclusion in a slightly different way: if there is a (partially) frozen half-BPS 7-brane other than the O7$^+$ plane plus D7-branes, 
there should also be a new, hitherto-unknown (partially) frozen half-BPS codimension-4 singularity in M-theory. The author considers this extremely unlikely.

F-theory has been used in various different constructions in the string theory literature.
Very often, it is implicitly assumed that the holomorphically varying axiodilaton corresponds to an elliptic fibration with a section and that 7-branes are not (partially) frozen, and it was not clear how serious the unintended consequences were.
In the last two years, genus-one fibrations without a section have been actively investigated, starting with \cite{Braun:2014oya}, but there are very few works on the O7$^+$ plane in the recent years, a notable exception being \cite{Garcia-Etxebarria:2013tba}.
It may be the time to start investigating F-theory setups with (partially) frozen singularities seriously.
The author hopes that this short note is useful as a first step in that direction, by showing that there is no other frozen 7-brane than the O7$^+$-plane.

\section*{Acknowledgements}
The author thanks K. Hori and D. R. Morrison for discussions,  M. Esole for valuable comments on an earlier draft of this note.
The author also thanks S.~Cherkis,  K.~Ohmori and A.~Tomasiello  for pointing out mistakes in the version 1 of the manuscript.
The work  is  supported in part by JSPS Grant-in-Aid for Scientific Research No. 25870159,
and in part by WPI Initiative, MEXT, Japan at IPMU, the University of Tokyo.

%\newpage

\bibliographystyle{ytphys}
%\baselineskip=.9\baselineskip
%\let\bbb\bibitem\def\bibitem{\itemsep1pt\bbb}
\bibliography{ref}

\providecommand{\href}[2]{#2}\begingroup\raggedright\begin{thebibliography}{10}

\bibitem{deBoer:2001px}
J.~de~Boer, R.~Dijkgraaf, K.~Hori, A.~Keurentjes, J.~Morgan, D.~R. Morrison,
  and S.~Sethi, ``{Triples, Fluxes, and Strings},'' {\em Adv. Theor. Math.
  Phys.} {\bfseries 4} (2002) 995--1186,
\href{http://arxiv.org/abs/hep-th/0103170}{{\ttfamily arXiv:hep-th/0103170}}.
%%CITATION = HEP-TH/0103170;%%.

\bibitem{Atiyah:2001qf}
M.~Atiyah and E.~Witten, ``{M Theory Dynamics on a Manifold of $G_2$
  Holonomy},'' {\em Adv. Theor. Math. Phys.} {\bfseries 6} (2003) 1--106,
\href{http://arxiv.org/abs/hep-th/0107177}{{\ttfamily arXiv:hep-th/0107177}}.
%%CITATION = HEP-TH/0107177;%%.

\bibitem{MorrisonTalk2002}
{\em {Half K3 Surfaces, or K3, G2, E8, M and all that}}.
\newblock Talk at Strings 2002.
\newblock \url{http://www.damtp.cam.ac.uk/events/strings02/avt/morrison/}.

\bibitem{Morrison:1996na}
D.~R. Morrison and C.~Vafa, ``{Compactifications of F-Theory on Calabi--Yau
  Threefolds -- I},''
  \href{http://dx.doi.org/10.1016/0550-3213(96)00242-8}{{\em Nucl. Phys.}
  {\bfseries B473} (1996) 74--92},
\href{http://arxiv.org/abs/hep-th/9602114}{{\ttfamily arXiv:hep-th/9602114}}.
%%CITATION = HEP-TH/9602114;%%.

\bibitem{Morrison:1996pp}
D.~R. Morrison and C.~Vafa, ``{Compactifications of F-Theory on Calabi--Yau
  Threefolds -- II},''
  \href{http://dx.doi.org/10.1016/0550-3213(96)00369-0}{{\em Nucl. Phys.}
  {\bfseries B476} (1996) 437--469},
\href{http://arxiv.org/abs/hep-th/9603161}{{\ttfamily arXiv:hep-th/9603161}}.
%%CITATION = HEP-TH/9603161;%%.

\bibitem{Witten:1997bs}
E.~Witten, ``{Toroidal Compactification without Vector Structure},'' {\em JHEP}
  {\bfseries 9802} (1998) 006,
\href{http://arxiv.org/abs/hep-th/9712028}{{\ttfamily arXiv:hep-th/9712028}}.
%%CITATION = HEP-TH/9712028;%%.

\bibitem{DelZotto:2014hpa}
M.~Del~Zotto, J.~J. Heckman, A.~Tomasiello, and C.~Vafa, ``{6D Conformal
  Matter},'' \href{http://dx.doi.org/10.1007/JHEP02(2015)054}{{\em JHEP}
  {\bfseries 02} (2015) 054},
\href{http://arxiv.org/abs/1407.6359}{{\ttfamily arXiv:1407.6359 [hep-th]}}.
%%CITATION = ARXIV:1407.6359;%%.

\bibitem{Ohmori:2015pua}
K.~Ohmori, H.~Shimizu, Y.~Tachikawa, and K.~Yonekura, ``{6D
  $\mathcal{N}{=}(1,0)$ Theories on $T^2$ and Class S Theories: Part I},''
  \href{http://dx.doi.org/10.1007/JHEP07(2015)014}{{\em JHEP} {\bfseries 07}
  (2015) 014},
\href{http://arxiv.org/abs/1503.06217}{{\ttfamily arXiv:1503.06217 [hep-th]}}.
%%CITATION = ARXIV:1503.06217;%%.

\bibitem{Witten:2000nv}
E.~Witten, ``{Supersymmetric Index in Four-Dimensional Gauge Theories},'' {\em
  Adv. Theor. Math. Phys.} {\bfseries 5} (2002) 841--907,
\href{http://arxiv.org/abs/hep-th/0006010}{{\ttfamily arXiv:hep-th/0006010}}.
%%CITATION = HEP-TH/0006010;%%.

\bibitem{Landsteiner:1997ei}
K.~Landsteiner and E.~Lopez, ``{New Curves from Branes},''
  \href{http://dx.doi.org/10.1016/S0550-3213(98)00022-4}{{\em Nucl. Phys.}
  {\bfseries B516} (1998) 273--296},
\href{http://arxiv.org/abs/hep-th/9708118}{{\ttfamily arXiv:hep-th/9708118}}.
%%CITATION = HEP-TH/9708118;%%.

\bibitem{Hanany:2001iy}
A.~Hanany and J.~Troost, ``{Orientifold Planes, Affine Algebras and Magnetic
  Monopoles},'' {\em JHEP} {\bfseries 08} (2001) 021,
\href{http://arxiv.org/abs/hep-th/0107153}{{\ttfamily arXiv:hep-th/0107153}}.
%%CITATION = HEP-TH/0107153;%%.

\bibitem{Schwarz:1995bj}
J.~H. Schwarz and A.~Sen, ``{Type IIA Dual of the Six-Dimensional CHL
  Compactification},''
  \href{http://dx.doi.org/10.1016/0370-2693(95)00952-H}{{\em Phys. Lett.}
  {\bfseries B357} (1995) 323--328},
\href{http://arxiv.org/abs/hep-th/9507027}{{\ttfamily arXiv:hep-th/9507027}}.
%%CITATION = HEP-TH/9507027;%%.

\bibitem{Witten:1997kz}
E.~Witten, ``{New ``Gauge'' Theories in Six Dimensions},'' {\em JHEP}
  {\bfseries 01} (1998) 001,
\href{http://arxiv.org/abs/hep-th/9710065}{{\ttfamily arXiv:hep-th/9710065}}.
%%CITATION = HEP-TH/9710065;%%.

\bibitem{Kachru:1997bz}
S.~Kachru, A.~Klemm, and Y.~Oz, ``{Calabi-Yau Duals for CHL Strings},''
  \href{http://dx.doi.org/10.1016/S0550-3213(98)00228-4}{{\em Nucl. Phys.}
  {\bfseries B521} (1998) 58--70},
\href{http://arxiv.org/abs/hep-th/9712035}{{\ttfamily arXiv:hep-th/9712035}}.
%%CITATION = HEP-TH/9712035;%%.

\bibitem{Braun:2014oya}
V.~Braun and D.~R. Morrison, ``{F-Theory on Genus-One Fibrations},''
  \href{http://dx.doi.org/10.1007/JHEP08(2014)132}{{\em JHEP} {\bfseries 08}
  (2014) 132},
\href{http://arxiv.org/abs/1401.7844}{{\ttfamily arXiv:1401.7844 [hep-th]}}.
%%CITATION = ARXIV:1401.7844;%%.

\bibitem{Garcia-Etxebarria:2013tba}
I.~Garc{\'\i}a-Etxebarria, B.~Heidenreich, and T.~Wrase, ``{New
  $\mathcal{N}{=}1$ dualities from orientifold transitions - Part II: String
  Theory},'' \href{http://dx.doi.org/10.1007/JHEP10(2013)006}{{\em JHEP}
  {\bfseries 10} (2013) 006},
\href{http://arxiv.org/abs/1307.1701}{{\ttfamily arXiv:1307.1701}}.
%%CITATION = ARXIV:1307.1701;%%.

\end{thebibliography}\endgroup
\end{document}